\documentclass[aps,twocolumn,english,superscriptaddress,citeautoscript,preprintnumbers,amsmath,amssymb,floatfix,footinbib]{revtex4-1}
\usepackage{graphicx}
\usepackage{graphics}               
\usepackage{subfigure}
\usepackage{bm}   
\usepackage{amssymb } 
\usepackage{array}       
\usepackage[breaklinks=true,colorlinks,citecolor=blue,linkcolor=blue,urlcolor=blue]{hyperref}
\usepackage{float}
\usepackage[table]{xcolor}
\usepackage{cancel}
\usepackage[normalem]{ulem}

\definecolor{Cyan}{rgb}{0.4,1,0.7}
\definecolor{LightCyan}{rgb}{0.8,1,0.85}

\begin{document}
	
	\title{Electronic structure, magnetic and transport properties of antiferromagnetic Weyl semimetal GdAlSi}
	
	\author{Antu Laha}\email[]{antulaha.physics@gmail.com}
	\affiliation{Department of Physics and Astronomy, Stony Brook University, Stony Brook, New York 11794-3800, USA}
	\affiliation{Condensed Matter Physics and Materials Science Division, Brookhaven National Laboratory, Upton, New York 11973-5000, USA}
        \author{Asish K. Kundu}\email[]{asishkumar2008@gmail.com}
	\affiliation{Condensed Matter Physics and Materials Science Division, Brookhaven National Laboratory, Upton, New York 11973-5000, USA}
	\author{Niraj Aryal}\email[]{naryal@bnl.gov}
	\affiliation{Condensed Matter Physics and Materials Science Division, Brookhaven National Laboratory, Upton, New York 11973-5000, USA}
    \author{Emil S. Bozin}
	\affiliation{Condensed Matter Physics and Materials Science Division, Brookhaven National Laboratory, Upton, New York 11973-5000, USA}	
    \author{Juntao Yao}
		\affiliation{Condensed Matter Physics and Materials Science Division, Brookhaven National Laboratory, Upton, New York 11973-5000, USA}
        \affiliation{Department of Materials Science and chemical Engineering, Stony Brook University, Stony Brook, New York 11794-3800, USA}
        \author{Sarah Paone}
		\affiliation{Department of Physics and Astronomy, Stony Brook University, Stony Brook, New York 11794-3800, USA}
	\affiliation{Condensed Matter Physics and Materials Science Division, Brookhaven National Laboratory, Upton, New York 11973-5000, USA}
 \author{Anil Rajapitamahuni}
	\affiliation{National Synchrotron Light Source II, Brookhaven National Laboratory, Upton, New York 11973-5000, USA}
 \author{Elio Vescovo}
	\affiliation{National Synchrotron Light Source II, Brookhaven National Laboratory, Upton, New York 11973-5000, USA}
        \author{Tonica Valla}
        \affiliation{Donostia International Physics Center, 20018 Donostia-San Sebastian, Spain}
         \author{Milinda Abeykoon}
	\affiliation{National Synchrotron Light Source II, Brookhaven National Laboratory, Upton, New York 11973-5000, USA}
        \author{Ran Jing}
	\affiliation{Condensed Matter Physics and Materials Science Division, Brookhaven National Laboratory, Upton, New York 11973-5000, USA}
        \author{Weiguo Yin}
	\affiliation{Condensed Matter Physics and Materials Science Division, Brookhaven National Laboratory, Upton, New York 11973-5000, USA}
     \author{Abhay N. Pasupathy}
	\affiliation{Condensed Matter Physics and Materials Science Division, Brookhaven National Laboratory, Upton, New York 11973-5000, USA}
        \affiliation{Department of Physics, Columbia University, New York, NY, 10027, USA}
        \author{Mengkun Liu}
	\affiliation{Department of Physics and Astronomy, Stony Brook University, Stony Brook, New York 11794-3800, USA}
 \affiliation{Condensed Matter Physics and Materials Science Division, Brookhaven National Laboratory, Upton, New York 11973-5000, USA}
 \affiliation{National Synchrotron Light Source II, Brookhaven National Laboratory, Upton, New York 11973-5000, USA}
	  \author{Qiang Li}\email[]{qiangli@bnl.gov}
	\affiliation{Department of Physics and Astronomy, Stony Brook University, Stony Brook, New York 11794-3800, USA}
	\affiliation{Condensed Matter Physics and Materials Science Division, Brookhaven National Laboratory, Upton, New York 11973-5000, USA}

\begin{abstract}
 We report the topological electronic structure, magnetic, and magnetotransport properties of a noncentrosymmetric compound GdAlSi. Magnetic susceptibility shows an antiferromagnetic transition at $T_\mathrm{N}$ = 32 K. In-plane isothermal magnetization exhibits an unusual hysteresis behavior at higher magnetic field, rather than near zero field. Moreover, the hysteresis behavior is asymmetric under positive and negative magnetic fields. First-principles calculations were performed on various magnetic configurations, revealing that the antiferromagnetic state is the ground state, and the spiral antiferromagnetic state is a close competing state. The calculations also reveal that GdAlSi hosts multiple Weyl points near the Fermi energy. The band structure measured by angle-resolved photoemission spectroscopy (ARPES) shows relatively good agreement with the theory, with the possibility of Weyl nodes slightly above the Fermi energy. Within the magnetic ordered state, we observe an exceptionally large anomalous Hall conductivity (AHC) of $\sim$ 1310 $\Omega^{-1}$cm$^{-1}$ at 2 K. Interestingly, the anomalous Hall effect persists up to room temperature with a significant value of AHC ($\sim$ 155 $\Omega^{-1}$cm$^{-1}$). Our analysis indicates that the large AHC originates from the Berry curvature associated with the multiple pairs of Weyl points near Fermi energy.
\end{abstract}
	
\maketitle
	
\section{Introduction}
Topological Weyl semimetals (WSMs) host emergent bulk Weyl fermions and surface Fermi arc states which give rise to robust transport signatures \cite{RevModPhys.90.015001,RevModPhys.88.021004,RevModPhys.93.025002,Rev_adv_phy_X,GdPtB_NatureMaterials_2016,CaCdSn_PRB_2020,YbCdGe_PRB_2019,YbCdSn_PRB_2020}. Magnetic WSMs where Weyl fermions are stabilized by the broken time-reversal symmetry are particularly appealing since they can produce tunable topological states under external magnetic fields. For example, the separation between the Weyl nodes can be tuned by external magnetic fields to enhance the net Berry flux and results in a large anomalous Hall effect \cite{PRL_2014,ZrTe5_Nature_phys_2016,GdPtBi_natphy_2016,Mn3Sn_NatureMaterials_2017,Science_2016_Mn3Ge}. On the other hand, the tunable control of the nontrivial real space spin textures can lead to a large topological Hall effect \cite{Gd2PdSi3_Science_2019,MnSi_PRL_2009,MnGe_PRL_2011,Mn3Sn_Nature_2015,Mn3Sn/Ge_PRL_2017,EuAgAs_PRB_2021}. Therefore, it is worthwhile to study the interplay between the electronic band topology in momentum space and the topology of spin texture in real space.
	
 Most of the Weyl semimetallic phases are formed by breaking either inversion symmetry or time-reversal symmetry. The Weyl semimetal phase that simultaneously breaks both inversion and time-reversal symmetry is relatively rare in real materials. The $R$Al$X$ ($R$ = rare earth, $X$ = Si, Ge) family of materials that belong to this double-symmetry breaking Weyl semimetals group, exhibit several intriguing topological properties as well as a rich variety of magnetic properties. While commensurate collinear magnetic order is vastly studied in Weyl semimetals, CeAlGe shows several incommensurate, square-coordinated multi-$\vec{k}$ magnetic phases and topological Hall effect \cite{CeAlGe_PRL_2020}. A helical ferrimagnetic order is observed in the Weyl semimetal NdAlSi, whose wavelength is linked to the nesting vector between two topologically non-trivial Fermi pockets \cite{NdAlSi_Nature_materials_2020}. The bond-oriented Dzyaloshinskii–Moriya interactions associated with Weyl exchange processes promote helical magnetism. In SmAlSi, the Weyl electrons take part in magnetic interactions and the Weyl-mediated indirect exchange coupling between $f$-electrons induces spiral magnetism \cite{SmAlSi_PRX_2023}. Interstingly, a special type of topological state, Kramers nodal lines, is observed in SmAlSi, where the doubly degenerate nodal lines connect time-reversal invariant momenta \cite{SmAlSi_Comn_phys_2023}. These materials possessing Kramers nodal lines lead to several exotic properties including monopole-like spin texture, and the quantized circular photogalvanic effect \cite{Ag2Se0.3Te0.7_Nat_mat_2018,RhSi_PRL_2017}. Furthermore, the magnetic field-induced Lifshitz transition, van-Hove singularity, and large anisotropic magnetocaloric effect are observed in PrAlSi \cite{PrAlSi_npj_2023,PrAlSi_JAP_2020}.
 	
Among others, spiral magnetism is an intriguing phenomenon in Weyl semimetal state. In order to form a spiral magnetic order state, a very weak magneto-crystalline anisotropy is required, which allows the spins to arrange in spiral order. GdAlSi has a very weak magneto-crystalline anisotropy, suggesting it as a good candidate for spiral magnetism. Here, we report the electronic structure and physical properties of GdAlSi single crystals. GdAlSi crystallizes in a LaPtSi-type body-centered tetragonal structure with the polar space group $I4_1md$. A Weyl semimetallic state with multiple pairs of Weyl nodes is confirmed by the first principle calculations and angle-resolved photoemission spectroscopy (ARPES) measurement. Magnetic susceptibility shows an antiferromagnetic transition at $T_\mathrm{N}$ = 32 K. An unusual asymmetric hysteresis behavior is observed in the in-plane isothermal magnetization. Moreover, we observe Berry curvature induced large anomalous Hall effect in the antiferromagnetic ordered state as well as in the non-magnetic state.

\section{Methods}
\subsection{Experimental details}
\begin{figure}
		\centering
		\includegraphics[scale=0.85]{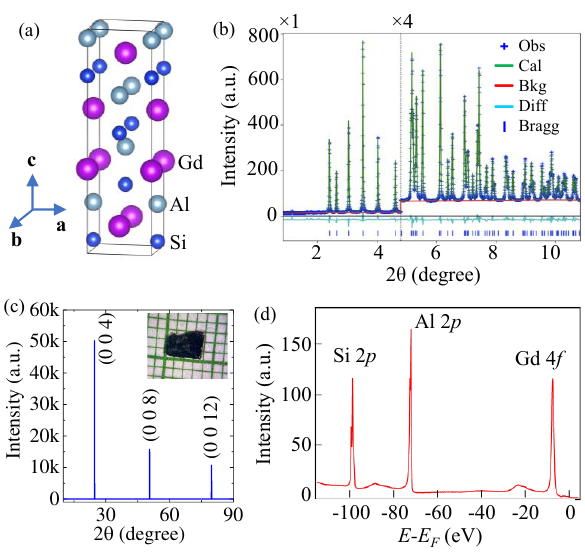} 
		\caption{ (a) Crystal structure of GdAlSi. (b) Powder x-ray diffraction pattern of crushed single crystals, recorded at room temperature (wavelength $\lambda$= 0.1665\AA). The intensity data for $2\theta \geq 4.8$ degree are multiplied by a factor 4 for clear visualization of the refinement. The observed intensity (blue scattered points), Rietveld refinement (solid green line), background (solid red line), difference between the experimentally observed and calculated intensities (solid cyan line), and Bragg peak positions (vertical blue bars) are shown. (c) Single crystal x-ray diffraction pattern (wavelength $\lambda$= 1.5406\AA). An image of a single crystal is shown in the inset (the grit scale: 1 mm). (d) The core level photoemission spectrum using a photon energy of $h\nu=$ 260 eV.}
		\label{Fig1}
	\end{figure}

GdAlSi single crystals were grown by the standard self flux method with excess Al as a flux \cite{Z_Fisk_Al_flux_growth,RAlGe_PRM_2019}. Gd ingot (99.9$\%$), Al ingot (99.999$\%$), and Si chips (99.999$\%$) in a molar ratio of 1:10:1 were mixed in an alumina crucible. The crucible was then sealed into a quartz tube under a partial pressure of argon gas. The content was heated to 1050$^\circ$C, kept for 24 hours at that temperature, and then cooled to 700$^\circ$C at a rate of 3$^\circ$C/hour. The excess flux could not be completely removed by centrifuging. Therefore, the remaining Al-flux was further removed by dissolution in a NaOH-H$_2$O solution. The very clean plate-like single crystals were filtered from the solution. The typical size of the crystal is 4mm$\times$3mm$\times$0.3mm as shown in the inset of Fig.~\ref{Fig1}(c). The crystal structure and phase purity were determined by x-ray diffraction and energy dispersive x-ray spectroscopy methods. The powder of crushed single crystals was characterized at 28-ID-1 beamline of the National Synchrotron Light Source II at Brookhaven National Laboratory with an x-ray wavelength of 0.1665\AA. The plane orientation of the single crystal was confirmed by Cu-K$_\alpha$ radiation in the Rigaku Smartlab diffractometer. Magnetotransport measurements were carried out in a physical property measurement system (PPMS, Quantum Design) via the standard six-probe method and magnetic measurements were executed in a magnetic property measurement system (MPMS, Quantum Design). In-situ strain study was performed in PPMS using Razorbill high-force strain cell. 
	
ARPES experiments were performed at the Electron Spectro Microscopy (ESM) 21-ID-1 beamline of the National Synchrotron Light Source II, USA. The beamline is equipped with a Scienta DA30 electron analyzer, with base pressure better than $\sim$ 1$\times$10$^{-11}$ mbar. Prior to the ARPES experiments, samples were cleaved using a post inside an ultra-high vacuum chamber (UHV) at $\sim$ 15 K. The total energy and angular resolution were $\sim$ 15 meV and $\sim$ 0.1$^\circ$, respectively. All measurements were performed using linearly-horizontal (LH) polarized light. The incident angle of light was 55 degrees with respect to the sample normal. The analyzer slits were oriented along the vertical direction. 

	\begin{figure*}
		\centering
		\includegraphics[scale=0.85]{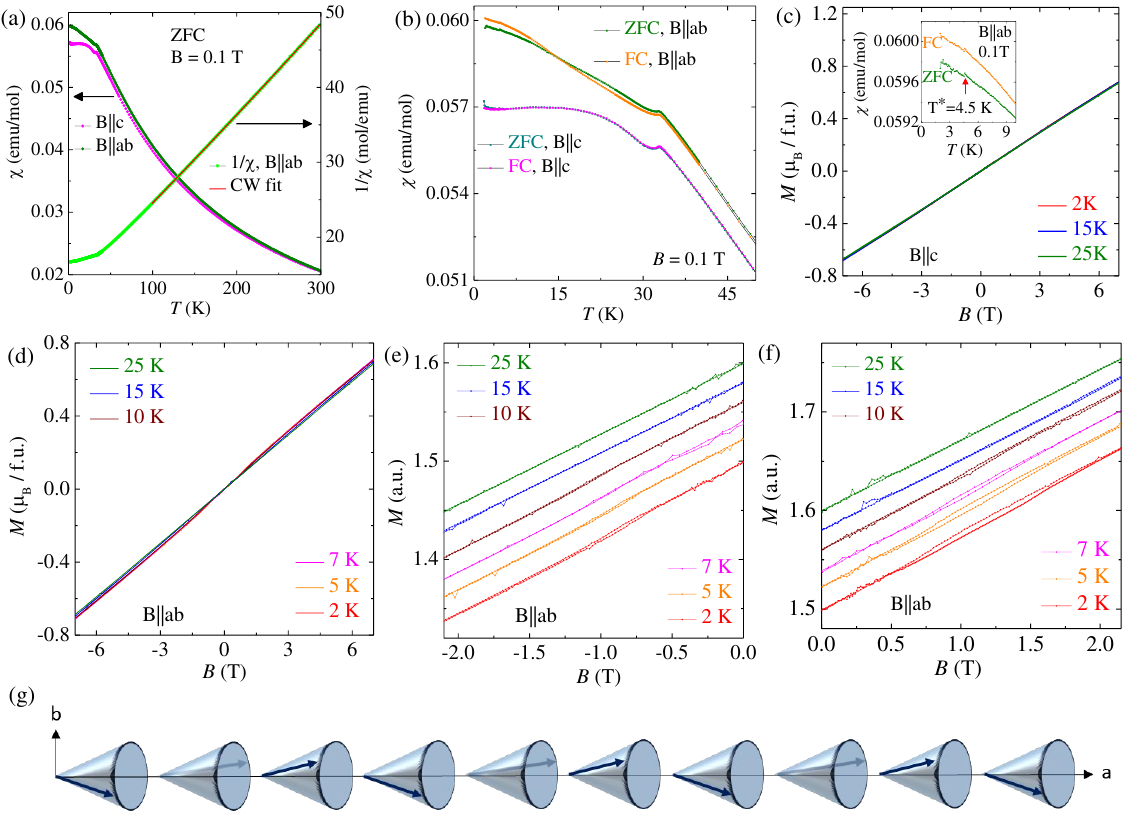} 
		\caption{(a) Temperature-dependent magnetic susceptibility in zero-field-cooled (ZFC) condition and the Curie-Weiss fit to the inverse magnetic susceptibility in the temperature range 100 K $-$ 300 K. (b) Magnetic susceptibility vs temperature in the temperature range 2 K - 50 K.(c) Isothermal magnetization curves for $B\parallel c$ at various temperatures. The inset shows a small kink at $T^*$ = 4.5 K for $B\parallel ab$. (d) Isothermal magnetization curves for $B\parallel ab$ at several temperatures. (e, f) The magnetization curves are uniformly shifted along vertical axis for clear visualization of the hysteresis loops. The area of hysteresis loop around 1 T is larger than that of around $-$1 T. (g) Schematic diagram of spin orientation for a typical spiral magnetic structure.}
		\label{Fig2}
	\end{figure*}
 
\subsection{Computational details}
The density-functional-theory (DFT) calculations were done using VASP DFT package~\cite{VASP1,VASP2,PAW}. The primitive Brillouin zone was sampled with a regular $11 \times 11 \times 11$ mesh containing 126 irreducible $k$ points. Perdew-Burke-Ernzerhof (PBE) exchange-correlation functional~\cite{PBE} within the generalized gradient approximation (GGA) was used in all the calculations.  The GGA + $U_{\textrm{eff}}$ method ~\cite{LDAU_Dudarev} was used to handle the Gd-4\textit{f} orbitals.
$U_{\textrm{eff}}$ of 6 eV was chosen in our calculations~\cite{NdAlSi_Nature_materials_2020}; however, we have also verified that the results presented here remain robust for a large range of $U_{\textrm{eff}}$ values. The calculations for the non-magnetic phase were done within open-core approximation by freezing the Gd-4\textit{f} states. The spin-orbit coupling (SOC) was treated in the second variation method. The slab and Weyl points calculations were done by using  Wannier90+WannierTools 
software~\cite{Wannier902014,WannierTools} by taking 56$\times$56 Wannierised Hamiltonian. Gd 5$p$, Gd 4$d$ orbitals, Al $p$, and Si $p$ orbitals were used in the Wannierisation procedure to accurately reproduce the DFT bands in the energy window from $-2$ to 2 eV.

\section{Results}
\subsection{Crystal structure and sample characterization}
 GdAlSi crystallizes in a LaPtSi-type body centered tetragonal structure with the polar space group $I4_1md$ (No. 109) as shown in Fig.~~\ref{Fig1}(a). The measured lattice parameters from the XRD refinement fitting are $a = b$ = 4.12584(4)\AA $~$and $c$ = 14.42816(20)\AA $~$[Fig.~\ref{Fig1}(b)]. Fig.~\ref{Fig1}(c) shows the XRD diffraction pattern for $c$-axis which confirms that the plate-like surface is $ab$-plane. Fig.~\ref{Fig1}(d) represents the core level photoemission spectrum of GdAlSi crystal, which clearly manifests the characteristic peaks originating from
Si 2$p$, Al 2$p$, and Gd 4$f$ orbitals.

 	\begin{figure*}
		\centering
		\includegraphics[scale=1.05]{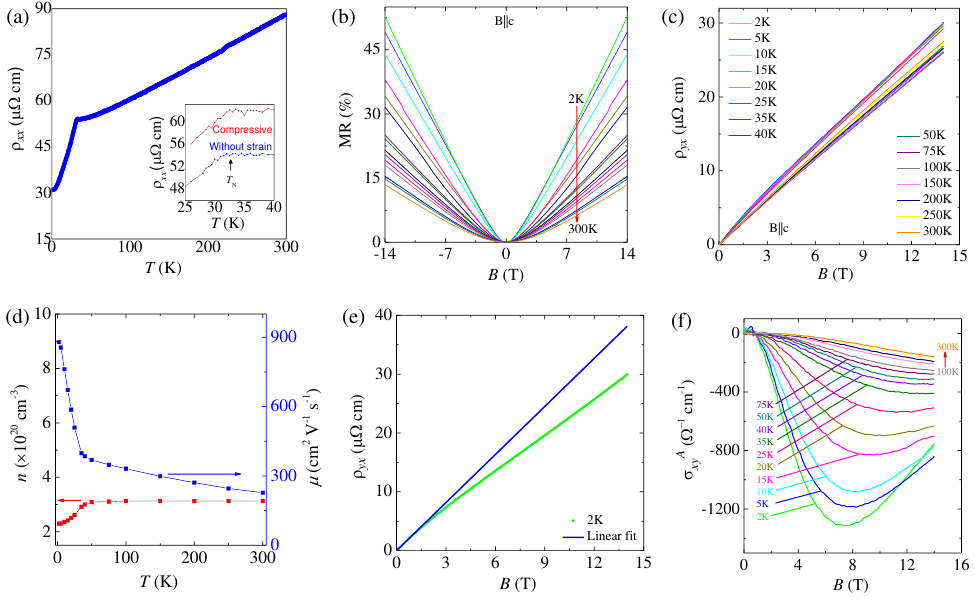} 
		\caption{(a) Electrical resistivity as a function of temperature. The inset shows a comparison of resistivity without strain and under compressive strain of 0.033\%. (b) Magnetic field dependence of transverse magnetoresistance (MR) at various temperatures from 2 K to 300 K. (c) Magnetic field dependence of Hall resistivity $\rho_{yx}$ in the temperature region 2 K - 300 K. (d) Temperature dependence of carrier density ($n$) and Hall mobility ($\mu$). (e) The $\rho_{yx}(B)$ at 2 K strongly deviates from linear field dependence. (f) Anomalous Hall conductivity ($\sigma_{xy}^A$) as a function of $B$ at various temperature from 2 K to 300 K.}
		\label{Fig3}
	\end{figure*}
 
\subsection{Magnetic properties}
Fig.~~\ref{Fig2}(a) shows the temperature dependent zero-field cooled (ZFC) magnetic susceptibility ($\chi$) along the crystallographic $c$-axis ($B\parallel c$) and in $ab$-plane ($B\parallel ab$). An antiferromagnetic (AFM) transition is observed at $T_\mathrm{N}$= 32 K due to the ordering of Gd$^{3+}$ moments. This compound shows a very weak magnetocrystalline anisotropy, $\chi_{c}/\chi_{ab}= 0.95$, which is similar to that of a spiral magnet SmAlSi \cite{SmAlSi_PRX_2023}. Such a weak magnetocrystalline anisotropy allows the spins to arrange
in a spiral magnetic order. To determine the effective magnetic moment of the Gd ions, the inverse susceptibility ($1/\chi$) is fitted with the modified Curie-Weiss law, $\chi(T)=\chi_0+C/(T-\theta_p)$, in the paramagnetic region (100 K - 300 K). Here, $\chi_0$, $C$, and $\theta_P$ are the temperature-independent susceptibility, Curie constant, and paramagnetic Curie temperature, respectively. The estimated effective magnetic moment of Gd$^{3+}$ is 8.11$\mu_B$ for $B\parallel ab$ and 8.26$\mu_B$ for $B\parallel c$ which are close to the theoretical value of $g\sqrt{S(S+1)}\mu_B=7.94\mu_B$ for $S=7/2$. The estimated values of the $\theta_P$ from the fitting are $-$109 K for $B\parallel ab$ and $-$116 K for $B\parallel c$. The negative values of paramagnetic Curie temperature are consistent with AFM ordering in this compound. We found that the $\chi(T)$ curve below $T_\mathrm{N}$ is quite different from that observed in a typical AFM system [Fig.~\ref{Fig2}(b)]. A bifurcation between ZFC and FC curves is observed below $T_\mathrm{N}$ for $B\parallel ab$ and these curves cross each other at 15 K. Moreover, isothermal M(B) curves show an unusual hysteresis behavior for $B\parallel ab$ as shown in Fig.~\ref{Fig2}(d),(e),(f). The hysteresis around 1 T and $-$1 T are not symmetric [ Fig.~\ref{Fig2}(e),(f)]. The area of hysteresis loop around 1 T is larger than that of around $-$1 T. We confirmed this unusual behavior by performing measurements repeatedly with several crystals. Overall, the hysteresis becomes weaker with increasing temperature, and it disappears for $T\geq$ 15 K [ Fig.~\ref{Fig2}(f)]. No hysteresis is observed for $B\parallel c$ as shown in Fig.~\ref{Fig2}(c). The value of magnetization of GdAlSi at 7 T is 0.7 $\mu_B$, which is an order of magnitude smaller than that of Gd$^{3+}$ moment. Similar behavior is also observed in its sister spiral magnetic compound SmAlSi \cite{SmAlSi_PRX_2023}. Moreover, SmAlSi also shows magnetic hysteresis only for $B\parallel a$ around 4 T rather than zero field region. A schematic diagram is shown in Fig.~\ref{Fig2}(g) for visualization of spin orientation in a typical spiral magnetic structure. Another small kink in $\chi(T)$ is observed at $T^*$= 4.5 K for $B\parallel ab$ as shown in the inset of Fig.~\ref{Fig2}(c). Further studies are required to get an insight into the magnetic ordered state of this compound. 
	
\subsection{Magnetotransport and in-plane strain effect}
The electrical resistivity ($\rho_{xx}$) is measured as a function of temperature within the $ab$-plane as shown in Fig.~\ref{Fig3}(a). The $\rho_{xx}(T)$ shows metallic behavior and a kink is observed at $T_\mathrm{N}$= 32 K due to the influence of magnetic ordering. We applied in-situ strain to the crystals along the in-plane direction while measuring temperature dependence of $\rho_{xx}$. The $\rho_{xx}$ increases with the increase of applied compressive strain. No changes in $T_\mathrm{N}$ are detected under maximum force of 39.2 Newton which corresponds to 0.033$\%$ compressive strain, indicating that the onset of magnetism is robust [the inset of Fig.~\ref{Fig3}(a)]. The estimated residual resistivity ratio [$\rho_{xx}(300K)/\rho_{xx}(2K)$=3] is comparable to that observed in other materials of this family \cite{PrAlSI_PRB_2020}. Fig.~\ref{Fig3}(b) displays the magnetic field dependence of transverse magnetoresistance (MR) for $B\parallel c$. The observed maximum MR is $53\%$ at 2 K and 14 T, which gradually decreases with increasing temperature. 
	
The field dependence of Hall resistivity, after removing the MR contribution using the expression $\rho_{yx}=[\rho_{yx}(B)-\rho_{yx}(-B)]/2$, is shown in Fig.~\ref{Fig3}(c). The positive $\rho_{yx}$ indicates that hole carriers dominate in GdAlSi. The carrier density and Hall mobility are estimated from the linear fitting of the $\rho_{yx}(B)$ curve at low field region using the relation $n$=1/($eR_\textrm{H}$) and $\mu$=$R_\textrm{H}$/$\rho_{xx}{(0)}$, where $R_\textrm{H}$ is the slope of $\rho_{yx}(B)$ curve. The temperature dependence of $n$ and $\mu$ are shown in Fig.~\ref{Fig3}(d). The carrier density of GdAlSi ($\sim 3\times 10^{20}$ cm$^{-3}$) is much smaller than that of typical metals ($10^{22}-10^{23}$ cm$^{-3}$) indicating a semimetallic behavior, consistent with the small value of density of states at the Fermi level ($E_F$) [see Fig.~\ref{fig:dft_nm}]. The values of $n$ and $\mu$ of GdAlSi are of the same order of magnitude as those reported for other members of this family \cite{CeAlSi_PRB_2021, PrAlSI_PRB_2020}. As shown in Fig.~\ref{Fig3}(d), above magnetic ordering temperature $T_N$, the carrier density is found to be constant as a function of temperature up to 300 K, suggesting nearly no changes in the electronic states near Fermi level, which is expected for a system like this with Fermi level substantially above band edge, and this is consistent with the DFT calculations discussed later. The Hall mobility decreases slowly with increasing temperature due to increased carrier scattering at higher temperature. Below $T_N$, magnetic order sets in and enhances as temperature decreases. Increased magnetic orders decreases magnetic scattering to the carriers that resulted in higher mobility and lower numbers of mobile carriers, i.e. lower value of $n$ at low temperatures.

The $\rho_{yx}(B)$ at 2 K exhibits an anomaly around 3 T and the curve strongly deviates from linear field dependence at higher magnetic field as shown in Fig.~\ref{Fig3}(e). The fitting with conventional two-band model fails to reproduce the Hall conductivity data. Rather, this behavior is reminiscent of the Berry curvature induced anomalous Hall effect (AHE) observed in AFM Weyl semimetals \cite{GdPtBi_natphy_2016,TbPtBi_PRB_2019}. To gain further insight, the anomalous Hall conductivity ($\sigma_{xy}^A$) is separated from the observed Hall conductivity ($\sigma_{xy}$) using the relation $\sigma_{xy}^A=\sigma_{xy}-\sigma_{xy}^N$, where $\sigma_{xy}^N$ is the normal Hall conductivity estimated from linear extrapolation of the low field fitting curve ($\rho_{yx}^N$). Both $\sigma_{xy}$ and $\sigma_{xy}^N$ are obtained through tensor conversions from the measured $\rho_{xx}$ and $\rho_{yx}$, $\sigma_{xy}=\rho_{yx}/(\rho_{xx}^2+\rho_{yx}^2)$ and $\sigma_{xy}^N=\rho_{yx}^N/(\rho_{xx}^2+\rho_{yx}^{N2})$. The field dependence of $\sigma_{xy}^A$ is plotted in Fig.~\ref{Fig3}(f) at various temperatures from 2 K to 300 K. The maximum value of $\sigma_{xy}^A$ reaches 1310 $\Omega^{-1}$cm$^{-1}$ around 7.5 T at 2 K. This value is quite large compared to those reported in RAlX family and other AFM Weyl semimetals [See Table \ref{table}]. The peak of $\sigma_{xy}^A$ becomes broader and shifts towards higher magnetic field with increasing temperature. Moreover, the value of $\sigma_{xy}^A$ decreases gradually with increasing temperature and reaches $\sim 155$ $\Omega^{-1}$cm$^{-1}$ at 300 K. We note that GdAlSi has the highest AHC, nearly two times higher than the second highest AHC material presented in table \ref{table}. Furthermore, the observed AHC in GdAlSi is present at room temperature, which is not observed in other materials compared in table \ref{table}.
	
	\begin{table}[h]
		\centering
		\caption {Anomalous Hall conductivity of RAlX family and other AFM Weyl semimetals}
		\renewcommand{\tablename}{Table.S\hskip-\the\fontdimen2\font\space}
		\label{table}
		\setlength\tabcolsep{1.5mm}
		\begin{tabular}{c c c c c c}
			\hline 
			Compound & $\sigma_{xy}^A$ ($\Omega^{-1}$cm$^{-1}$) & Ref. \\[1ex]
			\hline 
			NdAlGe & $\sim 430$ &  \cite{NdAlGe_PRB_2023}\\
			\hline
			CeAlSi & $\sim 550$ & \cite{CeAlSi_PRB_2023}\\
			\hline 
			PrAlGe & $\sim 680$ & \cite{PrAlGe_APL_2019} \\
			\hline  
			GdAlSi & $\sim 1310$ & [This work] \\
			\hline 
			\hline 
			GdPtBi & $\sim 30-200$ &  \cite{GdPtBi_natphy_2016}\\
			\hline
			TbPtBi & $\sim 744$ & \cite{TbPtBi_PRB_2019}\\
			\hline  
		\end{tabular}
	\end{table}

\begin{figure}
		  \includegraphics[width=0.45\textwidth]{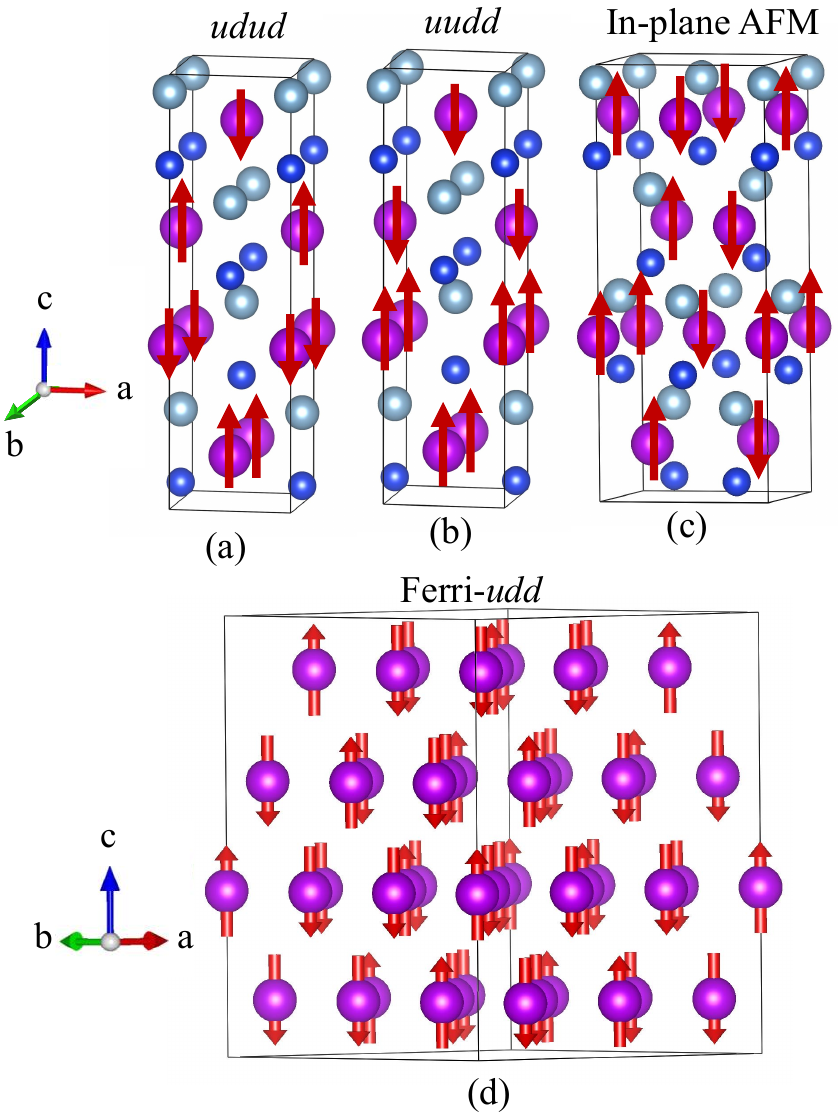}
		\caption{Different magnetic configurations used in the DFT calculations. The arrows represent the spin configurations of Gd-atoms. Only Gd-atoms are shown in panel (d) for clarity.}
		\label{fig:magnetic_conf}
\end{figure}

\begin{table}[t]
\vspace{-0.15in}
\begin{center}
\caption{\label{table:MagneticPatternEnergy}Calculated energy difference per formula unit in meV for different magnetic patterns (see text). 
}
\label{table:energetics}
\begin{tabular}{p{110pt}|p{30pt}p{55pt}}
\hline\hline
 Pattern & GGA & GGA+SO  \\ \hline
AFM-\textit{udud} & 0 & 0  \\ 
AFM-\textit{uudd}/\textit{uddu} & 10.6 & -  \\ 
FM & 14.2 & 13.2   \\ 
In-plane AFM  & 4.4 & 3.9  \\
Ferri-\textit{udd}  & 3.2  & -  \\
120$^{\circ}$ spiral  & 2.8 & 3.3  \\
 \hline\hline
\end{tabular}
\end{center}
\end{table} 

 \begin{figure*}
		\begin{center}
		  \includegraphics[width=0.98\textwidth]{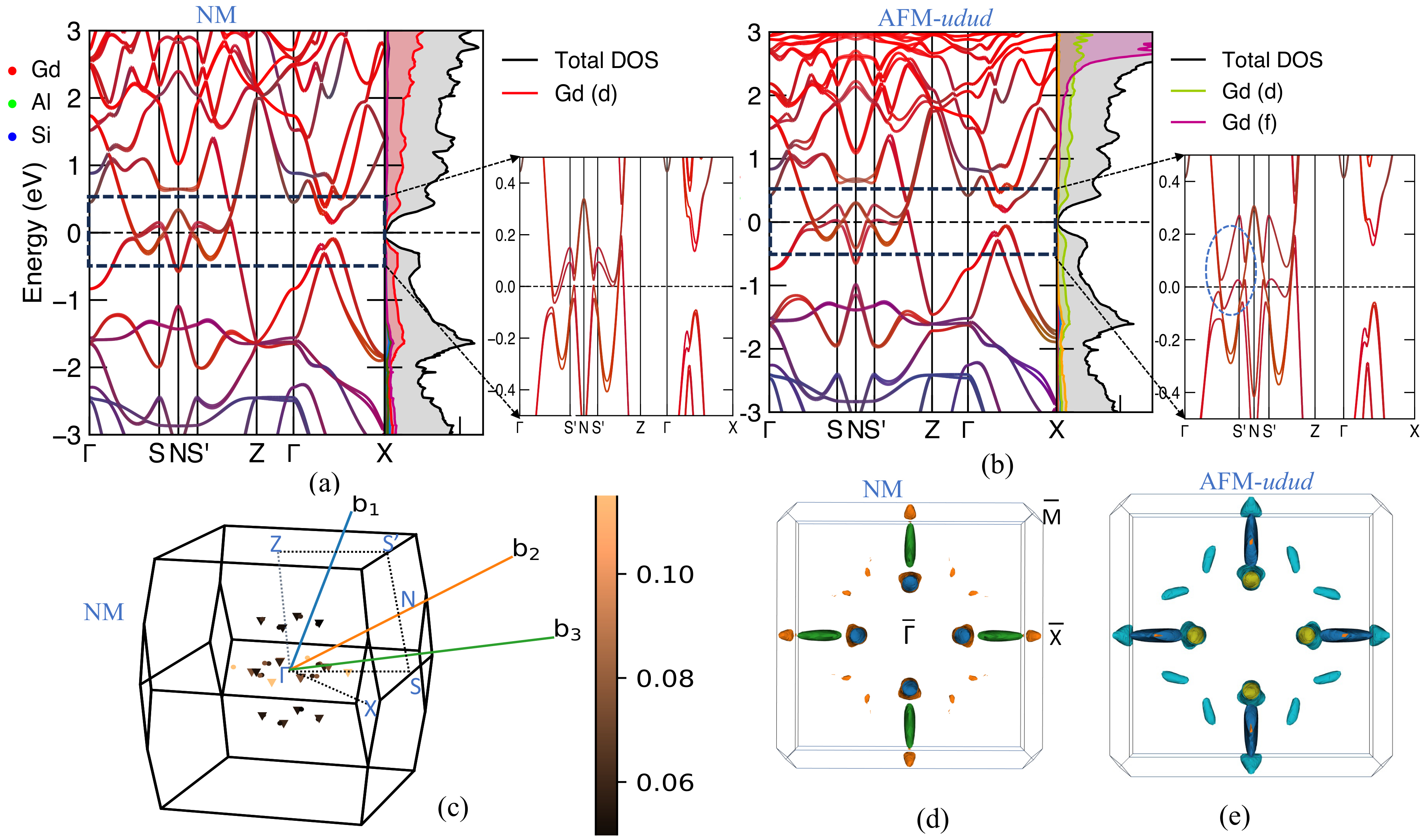}
        \end{center}
		\caption{Electronic structure of non-magnetic phase and lowest AFM phase of GdAlSi with the inclusion of SOC. (a, b) Band dispersion and orbital-resolved density of states for non-magnetic (NM) and \textit{udud} AFM phase. 
    The inset shows bands in a small energy window of $\pm$0.5 eV. (c) Distribution of Weyl points (WPs) in the primitive BZ of NM phase along with the high symmetry labels. The dots (triangles) show source (sink) of WPs. The color bar shows the location of WPs from $E_F$. (d, e) $k_z$-projected Fermi surface for NM and \textit{udud} AFM phase. The ``hammer''-like feature along $\Gamma$-S ($\bar{\Gamma}$-$\bar{\mathrm{X}}$ for $k_z$ projected FS) direction corresponds to the electron FS and small ellipsoids slightly off the $\Gamma$-X ($\bar{\Gamma}$-$\bar{\mathrm{M}}$) direction corresponds to the hole FS from Weyl bands.  See text for details.}
		\label{fig:dft_nm}
	\end{figure*}
  
\begin{figure*}
\begin{center}
\includegraphics[width=0.9\textwidth]{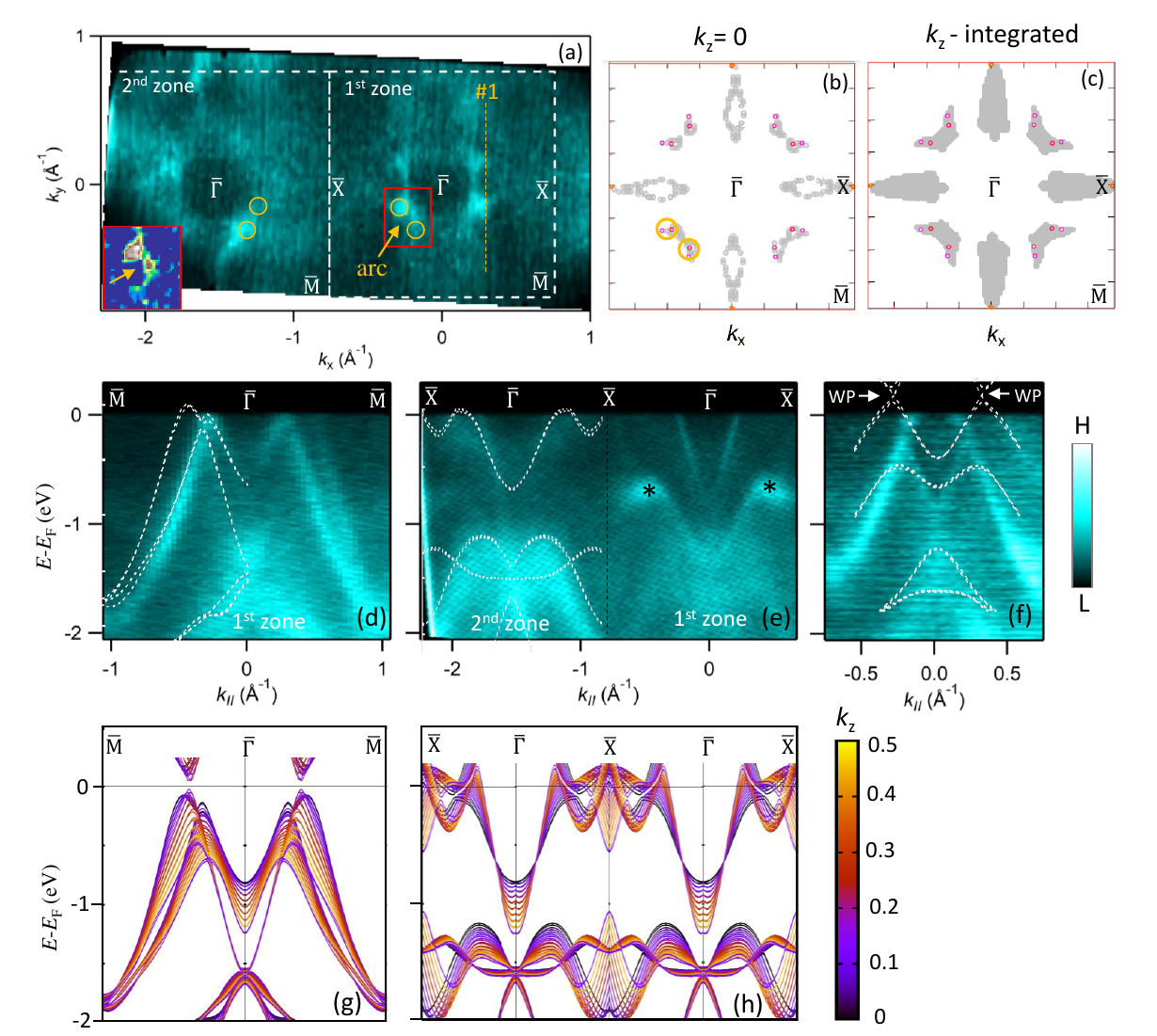}
\end{center}
\caption{Electronic structure of GdAlSi probed by ARPES. (a) Fermi surface map using a photon energy of 130 eV ($k_z\sim$ 0) and at $T$ = 15 K. The dashed squares outline the surface Brillouin zones. Circles illustrate the projection of Weyl cones. The inset displays the second derivative of the region enclosed by the red box, revealing a Weyl arc-like feature. (b) and (c), Theoretical constant energy contour at $E-E_F=-$120 meV in the $k_x-k_y$ plane for $k_z=$ 0 and $k_z$-integrated, respectively. The dots (magenta color) are the locations of Weyl points as obtained from calculations. (d) and (e), ARPES spectra along the $\bar{M}$-$\bar{\Gamma}$-$\bar{M}$, $\bar{X}$-$\bar{\Gamma}$-$\bar{X}$-$\bar{\Gamma}$-$\bar{X}$ directions of the SBZ. (f) ARPES spectra along the cut 1 as indicated in (a). Theoretical bands (white dashed lines) from the $k_z=$ 0 plane are superimposed on the ARPES spectra. (g) and (h), Theoretical bulk band structure integrated over $k_z$ from 0 ($\Gamma$) to 0.5 ($Z$) in the NM phase along the high symmetry directions $\bar{M}$-$\bar{\Gamma}$-$\bar{M}$ and $\bar{X}$-$\bar{\Gamma}$-$\bar{X}$-$\bar{\Gamma}$-$\bar{X}$, respectively.}
\label{fig6}
\end{figure*}
	
\subsection{DFT results}
 Fig.~\ref{fig:magnetic_conf} shows different magnetic patterns studied in this work using DFT calculation. Table.~\ref{table:MagneticPatternEnergy} shows the energy difference between various magnetic configurations. Our calculations find that all antiferromagnetic configurations are lower in energy compared to the ferromagnetic phase which is consistent with the experimental observation. In particular, A-type AFM pattern with in-plane FM and out-of-plane AFM arrangement of spins (\textit{udud}) [Fig.~\ref{fig:magnetic_conf}(a)] is found to be  lower in energy compared to up-down-down-up (equivalently up-up-down-down) [Fig.~\ref{fig:magnetic_conf}(b)] and in-plane AFM [Fig.~\ref{fig:magnetic_conf}(c)] arrangement. We also studied in-plane ferrimagnetic pattern~\cite{NdAlSi_Nature_materials_2020} by fixing spins in the up-down-down (\textit{udd}) direction along 110-axis [Fig.~\ref{fig:magnetic_conf}(d)] Similarly, $120^{\circ}$ rotated spiral spin arrangement along 110 direction was also investigated~\cite{SmAlSi_PRX_2023}. While both of these patterns were found to have slightly higher energy than the \textit{udud} pattern, the numbers are well within the uncertainty of the DFT calculations. The magneto-crystalline anisotropy energy is found to be weak in this system ($<$1 meV per Gd atom). This is consistent with the experimentally measured magnetic anisotropy for in-plane and out-of-plane applied magnetic field  [see Fig.~\ref{Fig2}(a)]. In addition to the GGA functional [Table.~\ref{table:energetics}], we also used SCAN-GGA functional to perform total energy calculations between different magnetic configurations for consistency check~\cite{SCAN2015,PhysRevB.108.054424}. SCAN functional also gave \textit{udud}  phase as the lowest energy state by similar order of magnitude.

 In Fig.~\ref{fig:dft_nm}, we compare the DFT calculated electronic structure of GdAlSi between the non-magnetic (NM) and lowest AFM phase (\textit{udud}) found from our fixed-spin DFT calculations. Since there are no significant differences in the electronic dispersion between the magnetic and NM phase [Fig.~\ref{fig:dft_nm}(a) and \ref{fig:dft_nm}(b)], we will first describe the electronic dispersion of NM phase in detail. The calculation for the NM phase was done within the open-core approximation where the Gd-4\textit{f} states were frozen. 
 Fig. ~\ref{fig:dft_nm}(a) shows atom-projected electronic bands and density of states which reveal that the states around the Fermi level are derived mainly from Gd-\textit{d} with small contributions from \textit{p}-states of Gd, Al, and Si. Linear bands exist in the vicinity of the Fermi level which is consistent with vanishing DOS. While there are SOC induced gaps along high symmetry lines, band crossings exist along non-high symmetry directions giving rise to many Weyl points. An exhaustive search of band crossings in the entire BZ finds 18 pairs of Weyl points with Chern number of $\pm1$ in the primitive BZ lying at around $\sim$ 50 to 100 meV above $E_F$. Despite having many pairs of Weyl nodes, most of them are in close vicinity to the Fermi level. While this system is very far from the ideal regime with single pairs of Weyl nodes, it still presents a unique opportunity to study Weyl mediated interactions due to the proximity of Weyl nodes to the Fermi level \cite{1,2,3,4,5,6,7}. The Fermi surface plot [Fig.~\ref{fig:dft_nm}(d)] shows ``hammer''-like feature from the electron bands along the $\bar{\Gamma}-\bar{X}$ direction and small ellipsoids from hole Weyl bands slightly away from $\bar{\Gamma}-\bar{M}$ direction. The electronic dispersion and FS plot for \textit{udud} AFM phase shown in Figs.~\ref{fig:dft_nm}(b) and ~\ref{fig:dft_nm}(e) are almost identical to the NM phase. The Gd-4\textit{f} states, which are at $-$6 eV below $E_F$ are pushed to $-$10 eV by the application of Hubbard $U$ of 6 eV. There is a weak hybridization between the Gd-\textit{5d} dominated Weyl bands with the Gd-4\textit{f} states which enhances the exchange splittings between the Kramer's pairs as highlighted by the dotted blue circle in the inset plot of Fig.~\ref{fig:dft_nm}(b). However, it is remarkable that except for a slight increment in the size of both the electron and hole pockets, the overall shape and topology of the FS remain intact. Therefore, in the next section, we use the electronic dispersion of the NM phase to make a comparison with the ARPES experiment. 

 \subsection{ARPES}
To experimentally probe the electronic structure of GdAlSi, we have performed ARPES measurements. Fig.~\ref{fig6}(a) represents the Fermi surface (FS) of GdAlSi measured in the AFM phase. We used a photon energy of 130 eV that covers two successive surface Brillouin zones (SBZs). One of the most notable features of the FS is the presence of high-intensity spots (enclosed by the circles) located near the $\bar{\Gamma}$ point, slightly offset along the $\bar{\Gamma}$-$\bar{M}$. A closed electron pocket is also observed around the $\bar{\Gamma}$. The overall features of the FS appear slightly different between the first and second zones, likely due to the photoemission matrix element effect. To understand the origin of the FS features, we have plotted the theoretical constant energy contour (at $E-E_F=-$120 meV) along with the Weyl nodes (magenta color dots) in the $k_x-k_y$ plane for $k_z=$ 0 and $k_z$ integrated from 0 ($\Gamma$) to 0.5 ($Z$) in Figs.~\ref{fig6}(b) and \ref{fig6}(c), respectively. Comparison of experimental and theoretical results suggests that the high-intensity spots and arc-like features [the inset of Fig.~\ref{fig6}(a)] observed in the ARPES Fermi surface possibly originate from the projection of the Weyl cones. Similar Weyl arcs were previously found in PrAlSi and SmAlSi \cite{SmAlSi_Comn_phys_2023,PrAlSi_SmAlSi_PRB_2023, PrAlGe_NatComn_2020}. ARPES spectra along the $\bar{\Gamma}$-$\bar{M}$ and $\bar{\Gamma}$-$\bar{X}$ directions of the SBZ are shown in Fig.~\ref{fig6}(d) and \ref{fig6}(e), respectively. The theoretical bands from the $k_z=$ 0 plane are overlayed on top of the ARPES spectra for comparison. A relatively good agreement between theory and experiment is obtained when the theoretical bands are shifted upward by $\sim$ 120 meV. This energy shift is most likely related to the surface effects. Overall, the ARPES spectra show some broadening due to the finite $k_z$ broadening. This can be better understood by comparing ARPES spectra to the projected DFT band structure results [Fig.~\ref{fig6}(g) and (h)]. Although the theoretical band dispersions agree well with the ARPES along the $\bar{\Gamma}$-$\bar{M}$ direction, some discrepancies can be seen along the $\bar{\Gamma}$-$\bar{X}$ direction, especially in the first SBZ. For example, the M-shaped band (marked by an asterisk symbol) observed in ARPES at around $-$0.8 eV is not present in the theoretical calculations, even in the $k_z$ projected spectra [Fig.~\ref{fig6}(h)]. Thus, this M-shaped band is possibly originating from the surface states. In Fig.~\ref{fig6}(f), the ARPES spectrum cutting through the Weyl nodes [cut 1 in Fig.~\ref{fig6}(a)] is displayed, with DFT bands superimposed. The comparison indicates that the material exhibits Weyl-like band dispersion near the Fermi energy. However, we could not resolve the Weyl points as they are expected to be above the Fermi energy from our DFT calculations.
 
 \section{Discussions and conclusions}
We now discuss viable reasons for observing a large anomalous Hall effect in GdAlSi. The conventional ferromagnetism induced anomalous Hall resistivity [$\rho_{yx}^A(B) \propto M(B)$] is not expected here because of the linear field dependence of $M(B)$. Topologically non-trivial spin texture in real space with finite scalar spin chirality [$\vec{S_1}.(\vec{S_2} \times \vec{S_3})$] generates a pseudo magnetic field, that gives rise to topological Hall effect. This can be a possible reason of large unconventional Hall contribution in GdAlSi. A large topological Hall effect has been reported in SmAlSi with a spiral magnetic order. GdAlSi also shows very weak magneto-crystalline anisotropy ($\chi_c/\chi_{ab} = $0.95) similar to that of SmAlSi, which allows the spins to arrange in a spiral magnetic order \cite{SmAlSi_PRX_2023}. However, the mechanism of non-trivial spin texture can not explain the anomalous Hall effect in GdAlSi above $T_\mathrm{N}$. The $\rho_{yx}$ even at 300 K still deviates from the linear field dependence. A significant $\sigma_{xy}^A$ ($\sim 155$ $\Omega^{-1}$cm$^{-1}$ at 14 T) is found at 300 K that indicates a different origin of AHE. In this context, we note that the band structure calculations reveal multiple Weyl points near Fermi energy and the Berry curvature associated with these Weyl points can produce a large anomalous Hall effect. 

It may be instructive to make a crude estimation of anomalous Hall conductivity arising from Berry curvature associated with the Weyl nodes, using the expression $\sigma_{xy}^A$ = $\frac{k}{\pi}.\frac{e^2}{h}$ \cite{AHC_PRL_2011,AHE_PRL_2014}. The estimated value of AHC for a single pair of Weyl nodes is $\sim 164$ $\Omega ^{-1}$cm$^{-1}$, where $k$ is the average distance (=0.13298 \AA$^{-1}$) between the pairs of Weyl points from our DFT calculation. We note that $\sigma_{xy}^A$ depends on the detailed band structure, in particular the orientation of the Weyl nodes and temperature. The total estimated value of $\sigma_{xy}^A$ including 18 pairs of Weyl nodes is $\sim 2900$ $\Omega ^{-1}$cm$^{-1}$. This very crude estimation is not far off from the experimental value $\sim 1310$ $\Omega ^{-1}$cm$^{-1}$ at 2 K. 

In conclusion, our magnetic susceptibility measurements revealed that GdAlSi has an antiferromagnetic ground state below $T_\mathrm{N}$ = 32 K. Isothermal magnetization curves show an unusual asymmetric hysteresis behavior, suggesting a possible spiral magnetic order in the $ab$-plane. Using first-principles calculations, we show that GdAlSi hosts an antiferromagnetic Weyl semimetal ground state and a spiral magnetic order state as a close competing state within the uncertainty of calculation. We also find multiple Weyl points near Fermi energy. The Weyl semimetallic features are further confirmed by ARPES measurements. The large observed anomalous Hall effect is contributed to the Berry curvature associated with the Weyl nodes near Fermi energy. Evaluating the topological Hall effect from real space spin texture requires detailed information of the magnetic order of the system, such as the possible spiral configuration strongly suggested by our experiments and DFT calculations. Unfortunately, Gd is a neutron absorber, thus it is challenging to probe the magnetic ordering in GdAlSi using neutron scattering approach. Other methods, such as resonant elastic X-ray scattering and spectroscopy TEM, might be able to shed light on the magnetic ordering in this class of materials.      
\section*{ACKNOWLEDGMENTS}
We would like to thank John Tranquada and Yue Cao for helpful discussions. The research at Brookhaven National Laboratory was supported by the U.S. Department of Energy, Office of Basic Energy Sciences, Contract No. DE-SC0012704.

%
	
\end{document}